# A Data-Driven Method to Map the Functional Organisation of Human Brain White Matter

Yifei Sun[1,2], James M. Shine[2,3], Robert D. Sanders[2,4], Robin F. H. Cash[5,6], Sharon L. Naismith[2,7], Fernando Calamante[1,2,8], Jinglei Lv[1,2,4].

1. School of Biomedical Engineering, The University of Sydney, Sydney, 2050, Australia.
2. Brain and Mind Centre, The University of Sydney, Sydney, 2050, Australia.
3. School of Medical Science, The University of Sydney, Sydney, 2050, Australia.
4. Sydney Medical School, The University of Sydney, Sydney, 2050, Australia.
5. Department of Biomedical Engineering, The University of Melbourne, Victoria, 3010, Australia.
6. Department of Psychiatry, The University of Melbourne, Victoria, 3010, Australia
7. School of Psychology, The University of Sydney, Sydney, 2050, Australia.
8. Sydney Imaging, The University of Sydney, Sydney, 2050, Australia.

## Abstract

The white matter of the brain is organised into axonal bundles that support long-range neural communication. Although diffusion MRI (dMRI) enables detailed mapping of these pathways through tractography, how white matter pathways directly facilitate large-scale neural synchronisation remains poorly understood. We developed a data-driven framework that integrates dMRI and functional MRI (fMRI) to model the dynamic coupling supported by white matter tracks. Specifically, we employed track dynamic functional connectivity (Track-DFC) to characterise functional coupling of remote grey matter connected by individual white matter tracks. Using independent component analysis followed by k-medoids clustering, we derived functionally-coherent clusters of white matter tracks from the Human Connectome Project young adult cohort. When applied to the HCP ageing cohort, these clusters exhibited widespread age-related declines in both functional coupling strength and temporal variability. Importantly, specific clusters encompassing pathways linking control, default mode, attention, and visual systems significantly mediated the relationship between age and cognitive performance. Together, these findings depict the functional organisation of white matter tracks and provide a powerful tool to study brain ageing and cognitive decline.

## 1. Introduction

Intricate brain function arises from vast numbers of neurons and their complex connectivity [1-3]. As the primary substrate of neuronal computation, grey matter (GM) has been extensively investigated for its role in brain function, with numerous atlases delineating functional segregation across the cerebral cortex and subcortical nuclei [4-7]. In contrast, the functional role of white matter (WM) has received comparatively less attention. WM serves as the indispensable "wiring" system, enabling communication between distributed neural regions. The involvement of WM in functional integration is supported by strong evidence in clinical studies. For example, in epilepsy patients undergoing corpus callosotomy (surgical severing of the corpus callosum), inter-hemispheric functional coupling is dramatically reduced, whereas intra-hemispheric connectivity is largely preserved [8].

Driven by modern connectome studies highlighting the strong coupling between structural connectivity and functional connectivity (FC) [9], novel approaches began to investigate the functional dynamics supported by WM tracks. In particular, track dynamic functional connectivity (Track-DFC), defined as the time-varying functional coupling between GM regions connected by individual WM tracks, has been shown to synchronise with task paradigms, providing direct evidence of WM-supported functional dynamics [10-12]. A related approach, track-weighted dynamic functional connectivity (TW-DFC) [13] further demonstrated the potential of WM track-based dynamics for functional parcellation within the corpus callosum, underscoring the rich functional features embedded in WM pathways. Together, these findings motivate a systematic investigation of whole-brain Track-DFC to uncover the functional organisation of WM at the track level. Building on this foundation, we aim to extend existing Track-DFC studies by leveraging a large, publicly available dataset to comprehensively characterise how track-level functional dynamics can be grouped into functionally meaningful subunits across the brain.

In parallel, WM degeneration is a well-established hallmark of ageing, encompassing volume loss, myelin disruption, and the accumulation of WM hyperintensities [14-16], and is closely associated with functional alterations and cognitive decline [17]. Despite this, how track level functional dynamics evolve during healthy ageing and contribute to age-related cognitive changes remains largely unexplored. Elucidating these relationships is essential for understanding how WM supports functional brain organisation and cognitive performance across the adult lifespan.

In this study, we leveraged high-quality diffusion magnetic resonance imaging (dMRI) and resting-state functional MRI (fMRI) data from the Human Connectome Project (HCP) [18, 19] to systematically investigate the functional parcellation of WM tracks based on Track-DFC. Temporally concatenated independent component analysis (ICA) followed by unsupervised clustering was used to group Track-DFC from 100 unrelated HCP young adults into functionally coherent clusters. We then examined age-related changes in functional coupling strength and temporal variability of these clusters using data from the HCP ageing cohort and assessed their associations with cognitive performance. Together, this integrative framework provides novel insight into the functional organisation of WM tracks and a powerful tool to study brain ageing and cognitive decline.

## 2. Methods

### 2.1 Dataset and Preprocessing

This study utilised data from the HCP, including the young adult (HCP-Y) [18] and ageing (HCP-A) [19] cohorts. The sample comprised 100 unrelated young adults and 708 healthy ageing adults with complete resting-state fMRI data and detailed age information. Participants in the HCP-Y cohort were aged 22–36 years (mean = 29.11), whereas those in the HCP-A cohort were aged 36–90 years (mean = 59.66). For each participant, all four minimally preprocessed resting-state fMRI scans [20] in MNI space were used. HCP-Y scans have 2 mm isotropic resolution with TR = 720ms, TE = 33.1ms, and 1200 time points per scan, while HCP-A scans have 2 mm isotropic resolution with TR = 800ms, TE = 37ms, and 478 time points per scan. All fMRI data were further spatially smoothed (6mm FWHM) and temporally band-pass filtered (0.01–0.1Hz) to improve signal-to-noise ratio while preserving meaningful spontaneous fluctuations.

### 2.2 Track Dynamic Functional Connectivity

To compute Track-DFC [10], we employed a tissue-unbiased brain template, generated using multimodal registration of T1-weighted, T2-weighted, fractional anisotropy, and mean diffusivity images from 50 HCP young adult (HCP-YA) subjects [12]. The associated whole-brain tractogram template (with 90,000 tracks) was generated using probabilistic fibre tracking and Anatomical Constrained Tractography in MRtrix. Using non-linear warping derived from T1-weighted images, the tractogram template was further mapped to standard MNI space to ensure alignment with HCP fMRI data. Importantly, streamlines in this template were optimised to terminate at the GM-WM boundary [12], enabling faithful extraction of fMRI signals associated with the cortical regions connected by each track.

An overview of the Track-DFC pipeline is illustrated in Figure 1a&b. For each streamline, the two endpoints were mapped to MNI voxel space to extract the corresponding fMRI time series, yielding a pair of endpoint signals per WM track. Track-DFC was then estimated using a sliding-window approach [21] (Figure 1b), in which a fixed-length temporal window was shifted across the time series with overlap. A window length of approximately 60s was used to ensure stable FC estimates [22], corresponding to 84 time points for the HCP-Y cohort (60.48s) and 75 time points for the HCP-A cohort (60.00s). Within each window, Pearson's correlation coefficient was computed between the two endpoint time series segments to quantify functional coupling. Sliding the window with a step size of one repetition time (1×TR) yielded time-resolved connectivity profiles that capture dynamic fluctuations in functional coupling along individual WM tracks.

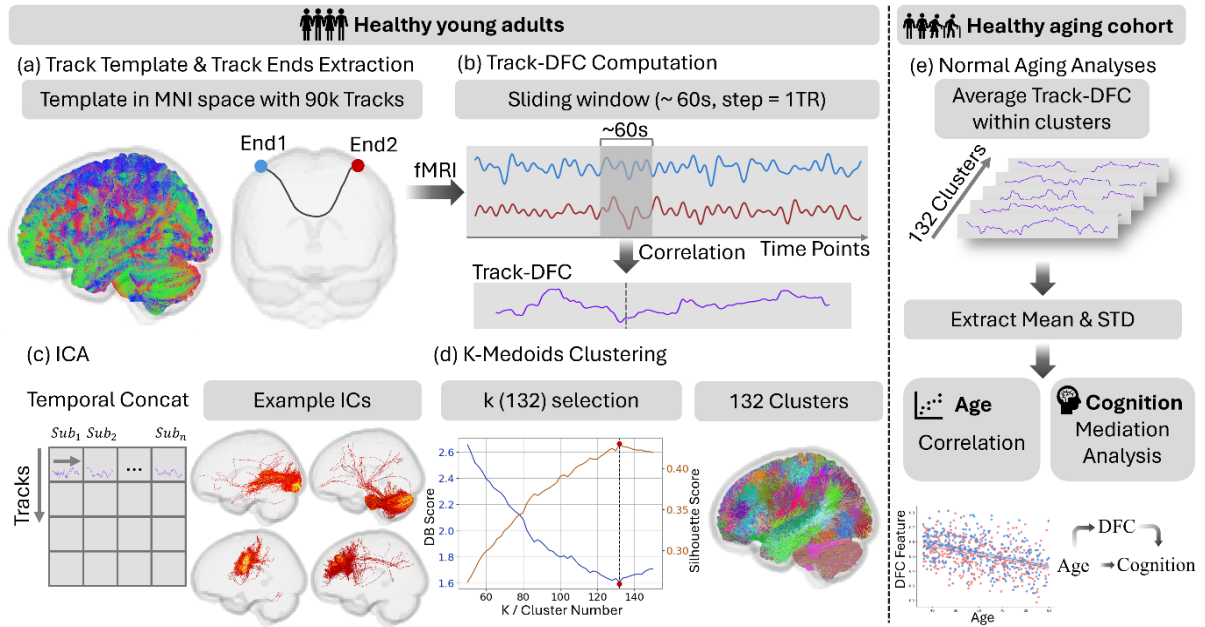

*Figure 1 Overview of the methods. (a) We used the template in MNI space with 90,000 white matter (WM) tracks. To calculate track dynamic functional connectivity (Track-DFC), we located the two endpoints of each WM track. (b) We extracted the corresponding time series at the two endpoints for each track from the resting-state functional magnetic resonance imaging (fMRI) data. Sliding-window method was applied to calculate the Track-DFC, with the window size of around 60s and a step size of one time point. Within each window, we calculated the Pearson's correlation, resulting in the final Track-DFC. For each subject, we calculated the 90k tracks' DFC and used them for the following steps. (c) 150 independent components (ICs) were obtained by using the group temporal concatenated melodic independent component analysis (ICA). Some example ICs are visualised to show the extracted WM track bundles located at different locations of the brain. (d) Based on the z-score maps from ICA, we used k-medoids clustering to generate the parcellation of WM tracks. K=132 was selected with the lowest Davis-Bouldin (DB) score and the highest Silhouette score within the test range of k, reflecting an optimal balance between cluster separability and the interpretability of the resulting clusters. Notably, steps (a) to (d) were all conducted based on the HCP healthy young adults. (e) We further explored the Track-DFC in the HCP healthy ageing cohort. Within each cluster, Track-DFC were first averaged across all tracks in the specific cluster to obtain the cluster-average Track-DFC, based on which the mean and standard deviation (STD) were extracted. We then explored the extracted cluster-level features' correlation with chronological age and their mediation effect in the age-cognition relationship, respectively.*

## 2.3 Dimension Reduction

We computed Track-DFC for 100 unrelated HCP-Y subjects and used group ICA analysis with temporally concatenation [23] to reduce dimensionality. ICA was performed using FSL MELODIC [24], with the Mixture Model-based inference threshold set to be 0.5. Our ICA decomposed the data into 150 independent components (ICs), balancing the need for principal patterns with the risk of over-fragmentation. Each resulting IC represented a spatial pattern of correlated resting-state brain activity, thereby reducing complexity and revealing independent DFC patterns in the dataset. Some example ICs are visualised in Figure 1c, showing WM bundles at different brain locations.

As ICA does not provide a deterministic parcellation of tracks – i.e. individual streamlines may contribute to multiple components, we used the z-score maps of ICs across the 90,000 tracks as feature representations for subsequent clustering. This enabled us to capture distributed functional contributions of WM tracks and leverage them as track-level features to group streamlines into functionally coherent clusters.

## 2.4 Clustering

For each WM track, we obtained a feature vector consisting of z-scores across ICs. We used the thresholded z-score maps provided by MELODIC ICA to reduce the impact of noise and improve clustering robustness. Furthermore, we excluded tracks with no variance in the thresholded z-score vector, leaving 82,103 candidate tracks with meaningful features for downstream clustering analysis.

Our goal was to group tracks into meaningful bundles based on IC loading patterns. Therefore, we employed k-medoids clustering, which is robust to noise and outliers [25], and supports flexible distance metrics. Cosine distance (Eq. 1 & Eq. 2) was used to quantify sample similarity, which is scale-invariant and emphasises vector orientation, making it well-suited for sparse data. This ensured clustering was driven by similarity in IC loadings patterns across tracks, rather than scale differences.

$$d_{Cosine}(x, y) = 1 - cosine\ similarity(x, y) \quad \text{Eq. 1}$$

$$cosine\ similarity\ (x, y) = \frac{x \cdot y}{||x|| * ||y||} \quad \text{Eq. 2}$$

To identify an optimal k balancing between the cluster quality and interpretability, we evaluated k ranging from 50 to 150 with increments of 2 and recorded inertia, Silhouette score [26], and Davies-Bouldin (DB) score [27]. The plot of inertia (elbow plot) did not show clear elbow point, so we primarily relied on other two metrics. The Silhouette score assesses how well each sample fits within its assigned cluster relative to others (higher the better), and our highest Silhouette score occurred at k = 132 within the tested range, as shown in Figure 1d. We further checked the DB score, which measures the ratio of within-cluster to between-cluster distances (lower the better), and observed the lowest DB score also at k = 132 (Figure 1d). As a result, we grouped our WM tracks into 132 clusters (Figure 1d).

## 2.5 Cluster Feature Extraction

We computed Track-DFC for HCP-A subjects using the same pipeline as for the HCP-Y. Using the track parcellation derived from HCP-Y, we extracted cluster-level DFC features for of the 132 clusters (Figure 1e). After averaging the DFC across tracks within each cluster, we computed the mean (which represents overall coupling strength) and standard deviation (STD, which represents DFC temporal variability [28]). Averaging all tracks' DFC within each cluster mitigated potential noise associated with single-track DFC estimates and reflects how synchronised or stable the cluster-level dynamics are. For each HCP-A subject, cluster DFC features were averaged across four scans.

## 2.6 Ageing Effect

We investigated ageing effects on the WM clusters by computing Spearman's correlations between chronological age and cluster-level DFC features (mean and STD). Correlation coefficients and false discovery rate (FDR)-corrected p-values were recorded. Clusters with statistically significant associations (corrected p < 0.05) were further ranked by their correlation coefficients.

## 2.7 Age-Cognition Mediation Analysis

We further investigated the relationships between WM Track-DFC features and the cognition using the HCP-A cohort. Given that cognitive performance is influenced by multiple factors, including age, education, etc., we conducted mediation analyses to test whether Track-DFC features mediate the association between age and cognition. We assumed that the cognitive measurements (dependent variable) can be directly affected by age (independent variable), and it may be indirectly affected by age through its effects on Track-DFC (mediator), as illustrated in Table 1. Sex and years of education were included as covariates, and one participant with missing education data was excluded.

The indirect (mediation) effects of DFC features were evaluated using bootstrap with 5000 iterations. We performed individual mediation analyses separately for each cluster and DFC feature, and the resulting p-values were FDR corrected across clusters. Clusters and DFC features showing significant mediation effect (corrected $p < 0.05$) were identified and ranked by mediation effect size.

We employed several commonly used cognitive measures available in HCP-A. The Montreal Cognitive Assessment (MoCA) total score [29] is a global cognitive measure ranging from 0 to 30, with higher scores indicating better performance across domains including memory, attention, language, and executive function. We also assessed Trail Making Test [30] A (TMT-A) and B (TMT-B) completion time (in seconds). In this test, TMT-A requires connecting numbers and reflects visual search and processing speed, whereas TMT-B requires alternating between numbers and letters and is more demanding for executive function. In addition, we investigated fluid, crystal, and total cognitive composite scores derived from NIH Toolbox Cognition tasks [31]. Fluid score reflects reasoning, processing speed, and executive function; crystallised composite score captures accumulated knowledge and language skills; and the total composite provides an overall index of global cognition performance.

For each cognitive measure, only subjects with available and valid scores were included. Specifically, one subject with a MoCA total score exceeding the valid range was excluded. Consistent with a prior study [32], we excluded one subject with a TMT-A completion time < 1s and two subjects with TMT-B completion time < 20s, and truncated the TMT-B time to 300s. The TMT-B minus TMT-A (TMT B-A) completion time was computed for subjects with valid completion time for both TMT-A and TMT-B. For the fluid composite score, 15 subjects with the value of 999 were excluded. After these quality-control procedures, the final HCP-A sample sizes were: MoCA, 706; TMT-A, 701; TMT-B, 702; TMT B-A, 698; Fluid composite, 604; Crystal composite, 603; and total composite, 603.

*Table 1 Regressions involved in the mediation analysis. Notably, sex and education were used as covariates in all regression listed here. DFC in this table denotes the cluster-level track dynamic functional connectivity (Track-DFC) feature. It is either cluster DFC mean or cluster DFC standard deviation. Cog denotes the specific cognitive measure.*

| Regression | Formula |
| --- | --- |
| DFC ~ Age | $DFC = b + b_1 Age + sex + education + e$ |
| Cog ~ DFC | $Cog = b + b_2 DFC + sex + education + e$ |
| Cog ~ Age | $Cog = b + b_3 Age + sex + education + e$ |
| Cog ~ Age + DFC | $Cog = b + b_5 Age + b_4 DFC + sex + education + e$ |
| Indirect | Effect = $b_1 * b_4$, Bootstrap n=5000 |

## 3. Results
### 3.1 Clusters of Tracks
Our method delivered 132 functional clusters of WM tracks, representing the segregation of functional roles associated with WM tracks. Figure 2a displays the clustered tractogram (top row), with each colour representing a functionally distinct cluster's WM track bundle. Four example clusters were visualised individually in the lower half of Figure 2a, with darker blue areas representing the high-density streamline core. Notably, these clusters were not constrained by anatomical boundaries but reflected shared functional signatures. The clustering revealed distinct bundles of streamlines, with groupings emerging in and between various brain regions.

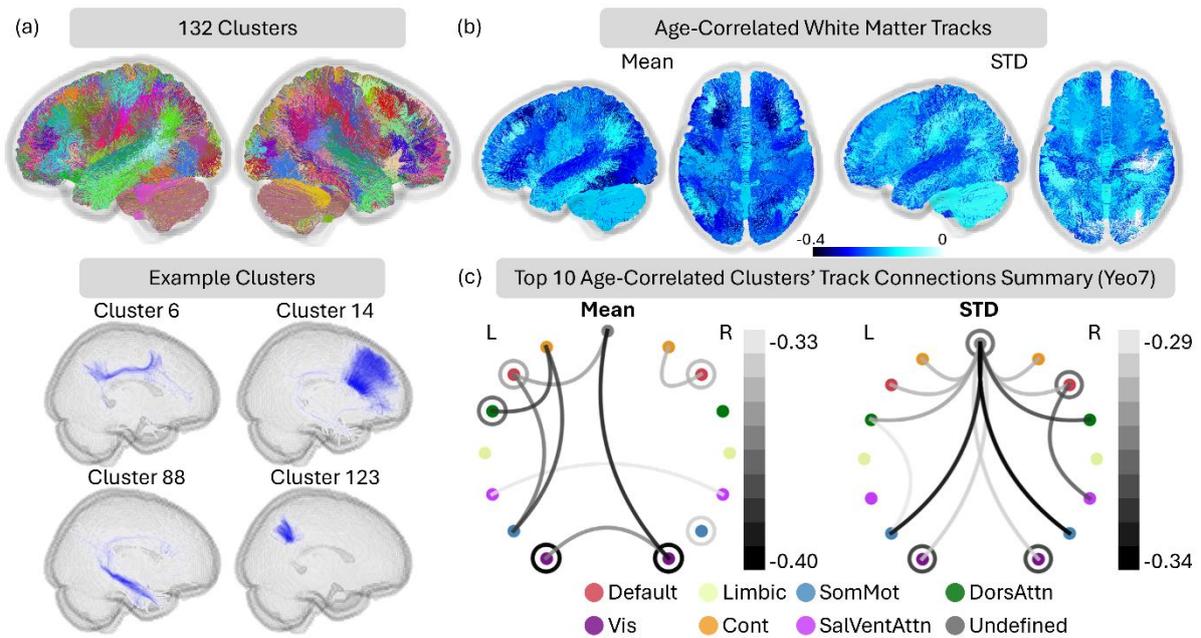

*Figure 2 (a) Top: colour-coded white matter track clusters generated from the 100 unrelated human connectome project healthy young subjects, showing 132 functionally grouped bundles with random colours assigned to distinct clusters. Bottom: Visualisation of four representative clusters with WM track bundles at different brain locations. Streamlines are rendered with high transparency so that darker blue regions indicate areas of higher streamline density, highlighting the core of the corresponding cluster. (b) Tracks with significant age correlation (false discovery rate (FDR)-corrected p < 0.05). Tracks are colour-coded corresponding to their correlation coefficient on cluster level. Darker blue colour represents stronger negative correlation, and brighter colour represents weaker correlation closer to 0. Both cluster-average mean and standard deviation showed negative correlation with age. (c) Track connection profiles for the top ten age-related clusters, characterised by major streamline endpoints labels in the Yeo7 atlas. Node colours indicate functional networks (Default – Default Mode, SomMot – Somatomotor, DorsAttn – Dorsal Attention, Vis – Visual, Cont – Control, SalVentAttn – Salience/Ventral Attention), with matching colours on the left and right sides representing the same network in the corresponding hemisphere. Arcs represent inter-network or interhemispheric intra-network links, while circles indicate intra-network connections within one hemisphere. Darker connections/circles show stronger age correlation.*

## 3.2 Effect of Age on Track-DFC Clusters

Our age correlation analyses identified statistically significant age-related WM track clusters. Out of the 132 clusters, 118 exhibited significant correlations ($p<0.05$, FDR corrected) between cluster-average Track-DFC mean and age, while 107 demonstrated significant correlations ($p<0.05$, FDR corrected) between DFC STD and age. These significant track clusters were visualised and colour coded by their correlation coefficients as shown in Figure 2b. Notably, both mean and STD of cluster-average Track-DFC showed negative correlations with age, indicating that dynamic connectivity strength and temporal variability decrease with age. To confirm the ageing effect on the cluster-level Track-DFC features, we visualised the scatter plot for the top three clusters as shown in Figure 3, with the linear fitting lines as the indicator for the trend. Similar significant negative trends were found in both male and female cohorts, represented by blue and red colour respectively.

We characterised cluster composition by identifying the functional networks (FNs) connected by their constituent WM tracks. Track endpoints were labelled using the Yeo 7-network atlas [4, 5], and cluster connection profiles were summarised by the frequency of network-to-network connection pairs. The seven networks include Visual (Vis), Somatomotor (SomMot), Dorsal Attention (DorsAttn), Salience/Ventral Attention (SalVentAttn), Limbic, Frontoparietal Control (Cont), and Default Mode (Default). For the most age-related clusters, dominant connections were visualised using connection diagrams (Figure 2c).

The top age-related clusters exhibited widespread connectivity spanning multiple cortical FNs. Clusters showing the strongest age effects on Track-DFC mean were characterised by a high proportion of cortical intra-network connections, particularly within the Vis, DorsAttn, Default, SalVentAttn, and SomMot networks. These clusters also included inter-network connections linking SomMot with Cont and Default networks, as well as Cont–DorsAttn and Cont–Default interactions. Collectively, these connections encompassed nearly all cortical FNs except the Limbic system, indicating a broad decline in functional coupling during normal ageing.

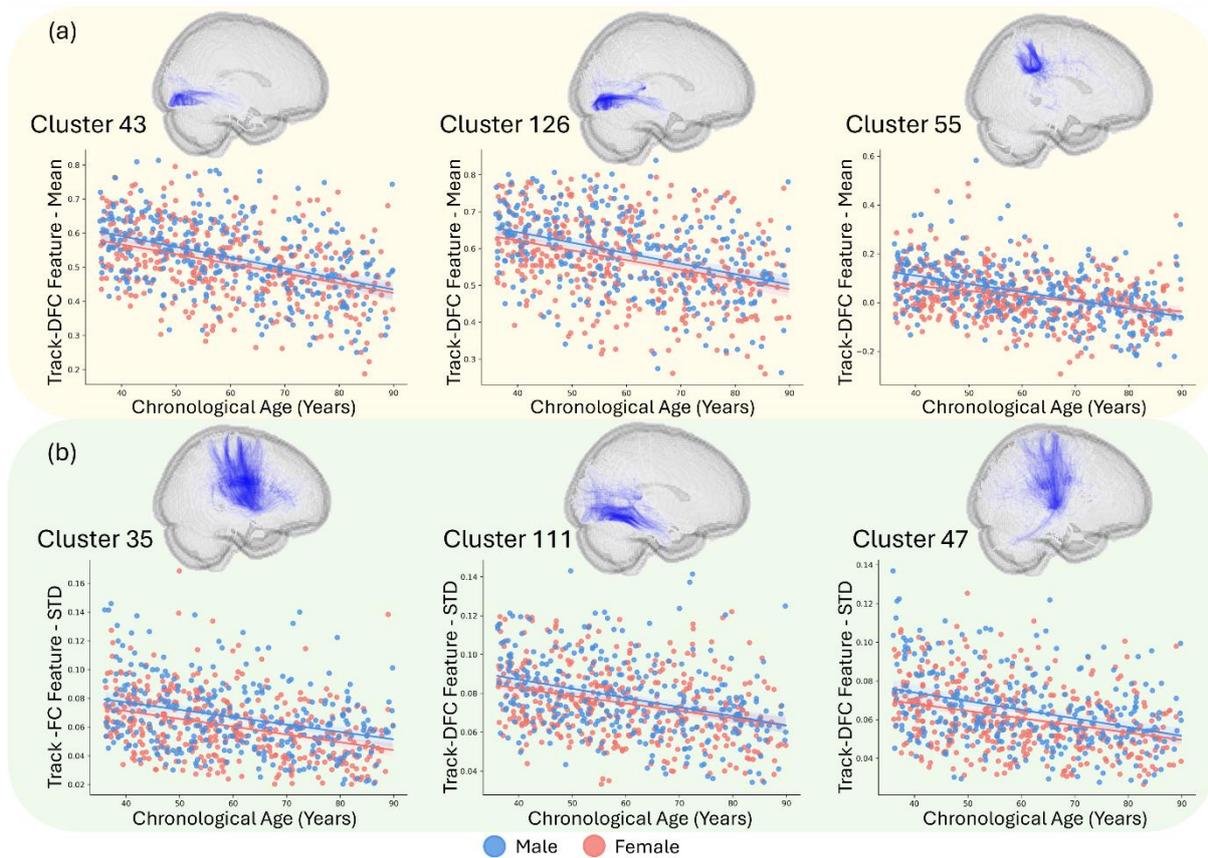

*Figure 3 Visualisation of the top age-sensitive white matter track clusters based on track dynamic functional connectivity (Track-DFC) metrics. (a) The three clusters with the strongest negative correlations between age and Track-DFC mean. (b) The three clusters most strongly correlated with age-related changes in Track-DFC standard deviation (STD). Each panel shows the spatial distribution of cluster tracks (top) and the corresponding scatter plot of the Track-DFC feature versus chronological age for male and female subjects (bottom), with fitted regression lines and confidence intervals of 95%. The linear fitting lines are just used as visual references to show the trend of the age-related changes.*

In contrast, age-related changes in cluster-average Track-DFC STD were most strongly associated with clusters involving tracks with one end in the SomMot, Default, Attention, Cont, and Vis networks. Beyond some inter-network connections, e.g. between Default and SalVentAttn networks and intra-Vis network connections, a substantial proportion of these connections extended from defined cortical FNs to regions not covered by the Yeo7 atlas, suggesting potential cortical-subcortical interactions. We further examined those endpoints' locations using the Freesurfer aseg atlas in MNI space (aseg_MNI152_2mm.nii.gz), which provides volumetric parcellations of subcortical structures derived from the Freesurfer average in MNI space. Within these top age-related clusters, the connected subcortical regions dominated in putamen and thalamus.

Together, these findings point to distinct mechanisms underlying age-related changes in WM supported functional dynamics: a widespread reduction in functional coupling strength across cortical networks, and a selective loss of temporal variability involving both cortical–subcortical and cortical-cortical interactions.

### 3.3 Mediation of Age and Cognition

Following FDR correction, only the cluster-average Track-DFC mean emerged as statistically significant mediator. Accordingly, this section focuses exclusively on the mediation results associated with this metric. Aside from TMT-A completion time and the crystal cognitive composite score, cluster-level Track-DFC mean showed significant mediation effects on other cognitive measures, including MoCA total score, TMT-B completion time, TMT completion time difference, fluid composite score, and the total composite score. To characterise connectivity patterns of clusters as significant mediators on cognitive measures, we visualised the endpoints connection profiles using Yeo7 atlas as shown in Figure 4. Four clusters with the strongest mediation effect were visualised in Figure 4.

Among cognitive measures, clusters with significant mediation effects exhibited diverse connectivity profiles. For the MoCA total score, 38 clusters demonstrated significant mediation effects, involving almost all FNs (Figure 4) except Limbic system. Clusters with the strongest effects were predominantly composed of tracks connecting Default, Cont, DorsAttn, and Vis networks, involving both intra-network and inter-network connections. Some connections anchored in SalVentAttn and SomMot systems were also observed.

For TMT-B completion time, we identified 17 clusters to be statistically significant in the mediation. These clusters primarily involved intra-network connections within Cont, Default, DorsAttn, and Vis networks and some inter-network bridges involving the Cont, Default, and DorsAttn systems. For the age-TMT time difference (TMT B-A) relationship, 16 clusters showed significant mediation effects with their Track-DFC mean. While the connectivity profile largely overlapped with the TMT-B results, some intra-network Cont connections and Cont-Default connections were no longer found to be significant mediators.

With regard to the fluid composite score, 15 clusters' Track-DFC mean demonstrated significant mediation effects. Clusters with the strongest mediation effect included connections within the Cont, Default, and DorsAttn networks. Additional connections were observed in the Vis, SalVentAttn, and SomMot systems. As a more general score, 17 clusters demonstrated significant mediation effects on the total cognitive composite score. These clusters largely overlapped with those identified for the fluid composite score and exhibited highly similar connectivity patterns.

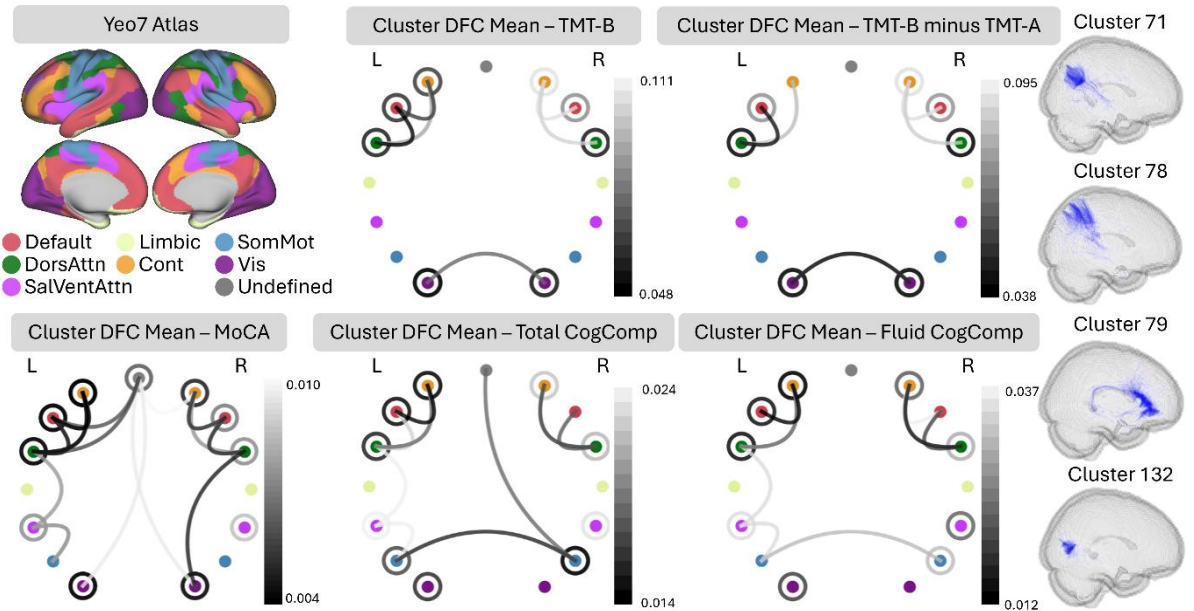

*Figure 4 Clusters showing statistically significant mediation effect on cognitive measures characterised by major streamline endpoints in the Yeo7 atlas (Top left, Default – Default Mode, SomMot – Somatomotor, DorsAttn – Dorsal Attention, Vis – Visual, Cont – Control, SalVentAttn – Salience/Ventral Attention). Each penal have the cluster-level feature and the cognitive measure specified on top of the connection plots. Node colours indicate functional networks, with matching colours on the left and right sides representing the same network in the corresponding hemisphere. Darker connections/circles correspond to stronger mediation effects (absolute value of effect). Arcs represent inter-network or interhemispheric intra-network links, while circles indicate intra-network connections within one hemisphere. Four clusters with the strongest mediation effects are visualised on the right. Streamlines are rendered with high transparency so that darker blue regions indicate areas of higher streamline density, highlighting the microstructural connection core of the corresponding cluster.*

# 4.  Discussion

## 4.1 Track-DFC and Clusters of Tracks

Our study introduces a track-based framework for characterising the functional organisation of WM through Track-DFC. By focusing on functional coupling between GM regions connected by WM pathways, this approach bridges structural connectivity with time-varying functional interactions and enables data-driven functional parcellation of WM independent of anatomical tract definitions.

This data-driven method allows for the characterisation of the human brain functional organisation supported by WM streamlines. In this work, we mainly utilised this method to map age-related changes in brain function and cognitive performance. Beyond ageing, this method could potentially be applied to study the network resilience and the effect of various diseases.

## 4.2 Age Effect on Track-DFC

Across the adult lifespan, we observed a widespread age-related decline in both the strength and temporal variability of Track-DFC, indicating a progressive weakening of long-range functional interactions supported by WM pathways. Reductions in Track-DFC mean likely reflect diminished efficacy of neural communication along WM streamlines. In parallel, reduced Track-DFC variability suggests a narrowing of the brain's dynamic range, potentially limiting its capacity for flexible network reconfiguration with advancing age.

The most direct interpretation links these functional decline to the known structural degradation of WM. Ageing is associated with demyelination, axonal loss and degeneration, and reduced neuroglial support [33-36]. These changes in WM can impair the efficiency and fidelity of signal transmission along the tracks. Furthermore, the DFC is a measure of the synchronisation of BOLD signals between the two WM track endpoints. Thus, age-related changes within the connected GM regions, such as reduced dendritic complexity [37] and altered neurotransmitter availability (e.g., dopamine, GABA) [38, 39], would also contribute to weaker and less dynamically sustained functional interactions.

Overall, our findings align with many fMRI studies reporting reduced static FC in the ageing brain [40, 41], but uniquely extend this to the level of functionally clustered WM pathways. The decreased Track-DFC mean and temporal variability serve as direct quantitative metrics of reduced efficacy and dynamics, respectively, within these critical long-range neural circuits.

*Clusters with high Age-Correlated Track-DFC Mean*

Clusters with strongest age-related declines in Track-DFC mean involved various cortical connections, including intra-network links within the Vis, Attention, Default, and SomMot networks, indicating a global vulnerability rather than a decline restricted to a single functional domain. The involvement of cross-hemisphere intra-network connections for Vis and SalVentAttn systems highlights the potential vulnerability of callosal pathways.

Beyond intra-network effects, we also observed connections linking SomMot to Cont and Default networks, as well as interconnections among Cont, Default, and DorsAttn systems. These higher-order networks are central to executive control [42], attentional regulation [43], and internally directed cognition [44], and thus reductions in their baseline DFC strength may underlie age-related declines in corresponding functions. Interestingly, reduced FC in higher-order networks, including SalVentAttn, Cont, and Default systems, has been reported in adults aged above 65 with postoperative delirium [45], raising the possibility that they may represent a shared locus of vulnerability across ageing and clinical populations.

Limbic network was found to be the least affected during normal ageing, although the anatomical regions comprising this network are known to exist at sites of poor image quality due to the overlying anatomical sinuses [46]. From a neurobiological perspective, this relative sparing may also reflect differential network vulnerability in healthy ageing, whereby higher-order association systems show earlier and more pronounced functional decline, while limbic circuits exhibit greater stability in absence of pathological conditions.

*Clusters with high Age-Correlated Track-DFC STD*

The reduction in DFC STD of cortical–subcortical pathways reflect diminished communication between cortical networks and subcortical hubs that regulate sensory, motor, and cognitive processes. The thalamus, in particular, is a critical relay centre for sensory, motor, and consciousness [47], while the putamen in basal ganglia mainly supports motor control and involves in cognition functions [48, 49]. Age-related decreases in the dynamics of these pathways could therefore underlie functional decline in both sensorimotor and cognitive domains. These findings extend the interpretation of ageing effects beyond cortical networks, suggesting that subcortical structures also play a role in the loss of functional dynamics.

We also observed Default-SalVentAttn connections in the top age-related clusters. Interestingly, the SalVentAttn network is thought to detect salient internal or external events and facilitate switching between the Default and Cont systems [50, 51]. Reduced temporal variability in this pathway during ageing may therefore reflect diminished efficiency of this switching mechanism, leading to less dynamic engagement and disengagement of the Default network in response to attentional demands. The decline in Track-DFC STD between SomMot and DorsAttn networks suggests a loss of dynamic variability in the intrinsic pathways that bridge spatial attention [43] and motor execution. Even at rest, this functional stability likely reflects a diminished capacity for the brain to rapidly integrate attentional cues with behavioural responses, potentially serving as a marker for age-related declines in processing speed.

### 4.3 Cognition-Relevant Clusters

Although both the mean and STD of Track-DFC declined with age, only Track-DFC mean of specific clusters acted as statistically significant mediators for age-cognition relationship, suggesting the importance of functional coupling strength in cognition.

Interestingly, we did not observe significant mediation effects of WM Track-DFC features on either TMT-A completion time or the crystal cognitive composite score. A potential reason is that TMT-A is a relatively low-demand task primarily indexing psychomotor speed and relatively simple visual scanning [30]. In turn, the crystallised cognition relies on long-term learning and knowledge accumulation, which are relatively stable across ageing [52, 53].

For the MoCA total score, we identified a relatively large number of clusters with significant mediation roles, suggesting that global cognitive status is supported by widespread connectivity across multiple networks. The dominance of Default, Cont, DorsAttn, Vis, and SalVentAttn networks highlights the integration of executive, attentional, and perceptual systems in supporting general cognition.

In age-TMT-B relationship, the clusters with significant mediation effect involves mainly the Default, Cont, DorsAttn, and Vis systems. This finding is consistent with the intense visual search, executive control, and attentional shifting demands of TMT-B task. To isolate executive demands, we explored the mediation effect on two TMT tasks time difference (TMT B-A). Our results highlighted the strongest mediation effect with DorsAttn connections, which are consistent with the role of the DorsAttn system in supporting goal-directed attention [43]. The persistence of mediation effects in DorsAttn connections after controlling for processing speed likely reflects that these pathways capture the core attentional demands of TMT-B. In contrast, the disappearance of intra-Cont and Cont-Default connections in the difference score suggests that these networks may contribute to common cognitive demands shared by both TMT-A and TMT-B. It is also worth noting that difference scores (e.g., TMT B-A) are known to have reduced reliability and provide conservative results [54], which may attenuate the statistical power to detect mediation effects in the difference score metric.

For the fluid and total composite score, beyond the commonly observed mediation effects involving Cont, Default, DorsAttn, and Vis connections, we identified connections within SomMot and SalVentAttn networks and some connections between SomMot and SalVentAttn networks for both fluid and total composite score. This suggests that the fluid cognition is mediated by both higher-order control and sensorimotor integration. Because the total composite score incorporates fluid abilities, it exhibited a highly similar mediation profile, indicating substantial overlap in the WM pathways through which ageing relates to overall cognitive performance.

Across these cognitive domains, significant mediation effects consistently involved clusters with WM connections in Cont, Default, DorsAttn, and Vis networks, highlighting a core set of large-scale systems through which age-related changes in Track-DFC may translate into cognitive decline.

### 4.4 Limitations and Future Directions

Despite its strengths, this study has several limitations. First, the use of template-based WM tracks may not fully capture individual anatomical variability. Although all analyses were performed in standard MNI space, this limitation may affect the precision of Track-DFC estimates, particularly in older adults with age-related structural changes. Second, our analyses were confined to a healthy ageing cohort, which limits generalisability to clinical populations and constrains interpretation to age–cognition mediation effects. Future work could incorporate subject-specific tractography to better account for individual differences and extend the framework to clinical and developmental cohorts across the lifespan. In addition, longitudinal designs will be essential for clarifying the temporal evolution of WM-supported functional dynamics and for establishing Track-DFC as a biomarker of brain health.

## 5. Conclusion

In this study, we introduced a novel framework for analysing and interpreting WM Track-DFC at resting-state, enabling the integration of structural pathways with time-varying functional coupling. This approach advances neuroimaging analysis methodology by moving beyond conventional region- or voxel-level analyses to capture connectivity features supported by anatomically defined WM tracks. Based on resting-state fMRI, we generated the functional clusters of WM tracks over the whole brain, establishing Track-DFC as a powerful methodology for probing the structural–functional interface. In healthy ageing cohort, we demonstrated that both Track-DFC mean and STD decline with age, reflecting reduced connectivity strength and narrowed dynamic range during normal ageing process. Importantly, the cluster-average Track-DFC mean mediated the relationship between age and certain cognitive measures, with effects concentrated in specific FNs. Therefore, our work provides new insight into the mechanisms of cognitive ageing and lays the foundation for future studies to explore more complex dynamic features and their roles throughout the lifespan and in clinical populations.

# Acknowledgement

J. Lv is supported by Brain and Mind Centre Research Development Grant, USYD-Fudan Brain and Intelligence Science Alliance Flagship Research Program, Moyira Elizabeth Vine Fund for Research into Schizophrenia Program. J. Lv, F. Calamante and S.L. Naismith are supported by an Australian Research Council Discovery Project (DP240102161). J.M. Shine is supported by National Health and Medical Research Council Grant (2016866), Australian Research Council Discovery Grant (DP240101295), and Australian Research Council Discovery Grant (DP250102186). R.F.H.C. is supported by the Australian National Health and Medical Research Council (Emerging Leadership Investigator grant no. 2017527).

Data were provided [in part] by the Human Connectome Project, WU-Minn Consortium (Principal Investigators: David Van Essen and Kamil Ugurbil; 1U54MH091657) funded by the 16 NIH Institutes and Centres that support the NIH Blueprint for Neuroscience Research; and by the McDonnell Centre for Systems Neuroscience at Washington University.

Research reported in this publication was supported by the National Institute On Ageing of the National Institutes of Health under Award Number U01AG052564. The content is solely the responsibility of the authors and does not necessarily represent the official views of the National Institutes of Health. The HCP-Ageing 2.0 Release data used in this report came from DOI: [10.15154/1520707](10.15154/1520707).

# References

[1] C. J. Honey *et al.*, "Predicting human resting-state functional connectivity from structural connectivity," *Proceedings of the National Academy of Sciences,* vol. 106, no. 6, pp. 2035-2040, 2009.

[2] A. Messé, D. Rudrauf, H. Benali, and G. Marrelec, "Relating structure and function in the human brain: relative contributions of anatomy, stationary dynamics, and non-stationarities," *PLoS computational biology,* vol. 10, no. 3, p. e1003530, 2014.

[3] E. Bullmore and O. Sporns, "Complex brain networks: graph theoretical analysis of structural and functional systems," *Nature reviews neuroscience,* vol. 10, no. 3, pp. 186-198, 2009.

[4] A. Schaefer *et al.*, "Local-global parcellation of the human cerebral cortex from intrinsic functional connectivity MRI," *Cerebral cortex,* vol. 28, no. 9, pp. 3095-3114, 2018, doi: 10.1093/cercor/bhx179.

[5] B. Thomas Yeo *et al.*, "The organization of the human cerebral cortex estimated by intrinsic functional connectivity," *Journal of neurophysiology,* vol. 106, no. 3, pp. 1125-1165, 2011, doi: 10.1152/jn.00338.2011.

[6] Y. Tian, D. S. Margulies, M. Breakspear, and A. Zalesky, "Topographic organization of the human subcortex unveiled with functional connectivity gradients," *Nature Neuroscience,* vol. 23, no. 11, pp. 1421-1432, 2020/11/01 2020, doi: 10.1038/s41593-020-00711-6.

[7] L. Fan *et al.*, "The Human Brainnetome Atlas: A New Brain Atlas Based on Connectional Architecture," (in eng), *Cereb Cortex,* vol. 26, no. 8, pp. 3508-26, Aug 2016, doi: 10.1093/cercor/bhw157.

[8] J. M. Johnston *et al.*, "Loss of resting interhemispheric functional connectivity after complete section of the corpus callosum," (in eng), *J Neurosci,* vol. 28, no. 25, pp. 6453-8, Jun 18 2008, doi: 10.1523/jneurosci.0573-08.2008.

[9] P. Fotiadis, L. Parkes, K. A. Davis, T. D. Satterthwaite, R. T. Shinohara, and D. S. Bassett, "Structure–function coupling in macroscale human brain networks," *Nature Reviews Neuroscience,* vol. 25, no. 10, pp. 688-704, 2024.

[10] J. Lv *et al.*, "Activated Fibers: Fiber-Centered Activation Detection in Task-Based FMRI," Berlin, Heidelberg, 2011: Springer Berlin Heidelberg, in Information Processing in Medical Imaging, pp. 574-587.

[11] J. Lv, M. Shine, F. Kong, and F. Calamante, "Mapping the Functional Role of White Matter Tracks by fusing Diffusion and Functional MRI," in *Proceedings of the international society for magnetic resonance in medicine*, 2024.

[12] J. Lv, R. Zeng, M. P. Ho, A. D'Souza, and F. Calamante, "Building a tissue-unbiased brain template of fiber orientation distribution and tractography with multimodal registration," *Magnetic Resonance in Medicine,* vol. 89, no. 3, pp. 1207-1220, 2023, doi: https://doi.org/10.1002/mrm.29496.

[13] F. Calamante, R. E. Smith, X. Liang, A. Zalesky, and A. Connelly, "Track-weighted dynamic functional connectivity (TW-dFC): a new method to study time-resolved functional connectivity," *Brain Structure and Function,* vol. 222, no. 8, pp. 3761-3774, 2017.

[14] J. Groh and M. Simons, "White matter aging and its impact on brain function," *Neuron,* vol. 113, no. 1, pp. 127-139, 2025.

[15]  M. Khodanovich *et al.*, "Age-Related Decline in Brain Myelination: Quantitative Macromolecular Proton Fraction Mapping, T2-FLAIR Hyperintensity Volume, and Anti-Myelin Antibodies Seven Years Apart," *Biomedicines,* vol. 12, no. 1, p. 61, 2024. [Online]. Available: https://www.mdpi.com/2227-9059/12/1/61.

[16]  F. A. S. de Kort *et al.*, "Cerebral white matter hyperintensity volumes: Normative age- and sex-specific values from 15 population-based cohorts comprising 14,876 individuals," *Neurobiology of Aging,* vol. 146, pp. 38-47, 2025/02/01/ 2025, doi: https://doi.org/10.1016/j.neurobiolaging.2024.11.006.

[17]  H. Liu *et al.*, "Aging of cerebral white matter," (in eng), *Ageing Res Rev,* vol. 34, pp. 64-76, Mar 2017, doi: 10.1016/j.arr.2016.11.006.

[18]  D. C. Van Essen *et al.*, "The WU-Minn human connectome project: an overview," *Neuroimage,* vol. 80, pp. 62-79, 2013, doi: 10.1016/j.neuroimage.2013.05.041.

[19]  S. Y. Bookheimer *et al.*, "The lifespan human connectome project in aging: an overview," *Neuroimage,* vol. 185, pp. 335-348, 2019.

[20]  M. F. Glasser *et al.*, "The minimal preprocessing pipelines for the Human Connectome Project," *Neuroimage,* vol. 80, pp. 105-124, 2013, doi: 10.1016/j.neuroimage.2013.04.127.

[21]  R. M. Hutchison *et al.*, "Dynamic functional connectivity: promise, issues, and interpretations," *Neuroimage,* vol. 80, pp. 360-378, 2013.

[22]  A. D. Savva, G. D. Mitsis, and G. K. Matsopoulos, "Assessment of dynamic functional connectivity in resting‐state fMRI using the sliding window technique," *Brain and behavior,* vol. 9, no. 4, p. e01255, 2019.

[23]  V. D. Calhoun, J. Liu, and T. Adalı, "A review of group ICA for fMRI data and ICA for joint inference of imaging, genetic, and ERP data," *Neuroimage,* vol. 45, no. 1, pp. S163-S172, 2009.

[24]  M. Jenkinson, C. F. Beckmann, T. E. Behrens, M. W. Woolrich, and S. M. Smith, "Fsl," *Neuroimage,* vol. 62, no. 2, pp. 782-790, 2012.

[25]  X. Jin and J. Han, "K-Medoids Clustering," in *Encyclopedia of Machine Learning*, C. Sammut and G. I. Webb Eds. Boston, MA: Springer US, 2010, pp. 564-565.

[26]  P. J. Rousseeuw, "Silhouettes: a graphical aid to the interpretation and validation of cluster analysis," *Journal of computational and applied mathematics,* vol. 20, pp. 53-65, 1987.

[27]  D. L. Davies and D. W. Bouldin, "A cluster separation measure," *IEEE transactions on pattern analysis and machine intelligence,* no. 2, pp. 224-227, 2009.

[28]  A. Elton and W. Gao, "Task‐related modulation of functional connectivity variability and its behavioral correlations," *Human brain mapping,* vol. 36, no. 8, pp. 3260-3272, 2015.

[29]  Z. S. Nasreddine *et al.*, "The Montreal Cognitive Assessment, MoCA: a brief screening tool for mild cognitive impairment," *Journal of the American Geriatrics Society,* vol. 53, no. 4, pp. 695-699, 2005.

[30]  R. M. Reitan, "Validity of the Trail Making Test as an indicator of organic brain damage," *Perceptual and motor skills,* vol. 8, no. 3, pp. 271-276, 1958.

[31]  N. Akshoomoff *et al.*, "VIII. NIH Toolbox Cognition Battery (CB): composite scores of crystallized, fluid, and overall cognition," *Monographs of the Society for Research in Child Development,* vol. 78, no. 4, pp. 119-132, 2013.

[32]  D. E. Cohen, H. Kim, A. Levine, D. P. Devanand, S. Lee, and T. E. Goldberg, "Effects of age on the relationship between sleep quality and cognitive

- [32] performance: Findings from the Human Connectome Project-Aging cohort," *International Psychogeriatrics,* vol. 36, no. 12, pp. 1171-1181, 2024/12/01/ 2024, doi: https://doi.org/10.1017/S1041610223000911.
- [33] E. Pannese, "Morphological changes in nerve cells during normal aging," *Brain Structure and Function,* vol. 216, no. 2, pp. 85-89, 2011/06/01 2011, doi: 10.1007/s00429-011-0308-y.
- [34] M. García-Domínguez, "Interplay Between Aging and Glial Cell Dysfunction: Implications for CNS Health," *Life,* vol. 15, no. 10, p. 1498, 2025.
- [35] A. Semyanov and A. Verkhratsky, "Chapter 4 - Neuroglia in aging," in *Handbook of Clinical Neurology*, vol. 209, A. Verkhratsky, L. D. de Witte, E. Aronica, and E. M. Hol Eds.: Elsevier, 2025, pp. 49-67.
- [36] A. Verkhratsky, "Neuroglial decline defines cognitive ageing," *Ageing and longevity,* vol. 6, no. 1, pp. 7-26, 2025.
- [37] M. E. Scheibel, R. D. Lindsay, U. Tomiyasu, and A. B. Scheibel, "Progressive dendritic changes in aging human cortex," *Experimental neurology,* vol. 47, no. 3, pp. 392-403, 1975.
- [38] M. D. Zuppichini *et al.*, "GABA levels decline with age: A longitudinal study," *Imaging Neuroscience,* vol. 2, pp. 1-15, 2024.
- [39] N. D. Volkow *et al.*, "Association between decline in brain dopamine activity with age and cognitive and motor impairment in healthy individuals," *American Journal of psychiatry,* vol. 155, no. 3, pp. 344-349, 1998.
- [40] J. R. Andrews-Hanna *et al.*, "Disruption of large-scale brain systems in advanced aging," *Neuron,* vol. 56, no. 5, pp. 924-935, 2007.
- [41] L. K. Ferreira and G. F. Busatto, "Resting-state functional connectivity in normal brain aging," *Neuroscience & Biobehavioral Reviews,* vol. 37, no. 3, pp. 384-400, 2013.
- [42] M. L. Dixon *et al.*, "Heterogeneity within the frontoparietal control network and its relationship to the default and dorsal attention networks," *Proceedings of the National Academy of Sciences,* vol. 115, no. 7, pp. E1598-E1607, 2018.
- [43] S. Vossel, J. J. Geng, and G. R. Fink, "Dorsal and ventral attention systems: distinct neural circuits but collaborative roles," (in eng), *Neuroscientist,* vol. 20, no. 2, pp. 150-9, Apr 2014, doi: 10.1177/1073858413494269.
- [44] R. L. Buckner, J. R. Andrews-Hanna, and D. L. Schacter, "The brain's default network: anatomy, function, and relevance to disease," (in eng), *Ann N Y Acad Sci,* vol. 1124, pp. 1-38, Mar 2008, doi: 10.1196/annals.1440.011.
- [45] N. L. Taylor *et al.*, "Dysfunctional resting state network connectivity predicts postoperative delirium after major surgery," *British Journal of Anaesthesia,* 2025/12/31/ 2025, doi: https://doi.org/10.1016/j.bja.2025.11.036.
- [46] D. Cordes, P. A. Turski, and J. A. Sorenson, "Compensation of susceptibility-induced signal loss in echo-planar imaging for functional applications☆," *Magnetic Resonance Imaging,* vol. 18, no. 9, pp. 1055-1068, 2000/11/01/ 2000, doi: https://doi.org/10.1016/S0730-725X(00)00199-5.
- [47] L. V. Marcuse *et al.*, "The thalamus: Structure, function, and neurotherapeutics," *Neurotherapeutics,* vol. 22, no. 2, p. e00550, 2025/03/01/ 2025, doi: https://doi.org/10.1016/j.neurot.2025.e00550.

[48] G. E. Alexander, M. R. DeLong, and P. L. Strick, "Parallel organization of functionally segregated circuits linking basal ganglia and cortex," *Annual review of neuroscience,* vol. 9, no. 1, pp. 357-381, 1986.

[49] S.-y. An, S.-H. Hwang, K. Lee, and H. F. Kim, "The primate putamen processes cognitive flexibility alongside the caudate and ventral striatum with similar speeds of updating values," *Progress in Neurobiology,* vol. 243, p. 102651, 2024/12/01/ 2024, doi: https://doi.org/10.1016/j.pneurobio.2024.102651.

[50] J. Schimmelpfennig, J. Topczewski, W. Zajkowski, and K. Jankowiak-Siuda, "The role of the salience network in cognitive and affective deficits," (in eng), *Front Hum Neurosci,* vol. 17, p. 1133367, 2023, doi: 10.3389/fnhum.2023.1133367.

[51] V. Menon and L. Q. Uddin, "Saliency, switching, attention and control: a network model of insula function," (in eng), *Brain Struct Funct,* vol. 214, no. 5-6, pp. 655-67, Jun 2010, doi: 10.1007/s00429-010-0262-0.

[52] C. N. Harada, M. C. Natelson Love, and K. L. Triebel, "Normal cognitive aging," (in eng), *Clin Geriatr Med,* vol. 29, no. 4, pp. 737-52, Nov 2013, doi: 10.1016/j.cger.2013.07.002.

[53] T. Salthouse, "Consequences of age-related cognitive declines," *Annual review of psychology,* vol. 63, no. 1, pp. 201-226, 2012.

[54] J. R. Edwards, "Ten difference score myths," *Organizational research methods,* vol. 4, no. 3, pp. 265-287, 2001.